\documentclass[12pt]{article}
\usepackage[english]{babel}
\usepackage{amsmath, amssymb}
\usepackage{graphicx}
\usepackage{natbib}

\newcommand{\mjup}{M_{\mathrm{J}}}
\newcommand{\miso}{m_{\mathrm{MISO}}}

\begin{document}

\title{Rendez-vous with massive interstellar objects, as triggers
of destabilisation}

\author{Denis V. Mikryukov$^{1}$ \& Ivan~I.~Shevchenko$^{1,2}$}

\maketitle

\noindent $^1$Saint Petersburg State University, 7/9
Universitetskaya nab., 199034 Saint Petersburg, Russia

\noindent $^2$Institute of Applied Astronomy, Russian Academy of
Sciences, 191187 Saint~Petersburg, Russia \\
Email: {\tt i.shevchenko@spbu.ru}

\abstract{We study how close passages of interstellar objects of
planetary and substellar masses may affect the immediate and
long-term dynamics of the Solar system. We consider two nominal
approach orbits, namely, the orbits of actual interstellar objects
1I/'Oumuamua and 2I/Borisov, assuming them to be typical or
representative for interstellar swarms of matter. Thus, the
nominal orbits of the interloper in our models cross the inner
part of the Solar system. Series of massive numerical experiments
are performed, in which the interloper's mass is varied with a
small step over a broad range. We find that, even if a Jovian-mass
interloper does not experience close encounters with the Solar
system planets (and this holds for our nominal orbits), our
planetary system can be destabilised on timescales as short as
several million years. In what concerns substellar-mass
interlopers (free-floating brown dwarfs), an immediate (on a
timescale of $\sim 10$--100~yr) consequence of such a MISO flyby
is a sharp increase in the orbital eccentricities and inclinations
of the outer planets. On an intermediate timescale ($\sim
10^3$--$10^5$~yr after the MISO flyby), Uranus or Neptune can be
ejected from the system, as a result of their mutual close
encounters and encounters with Saturn. On a secular timescale
($\sim 10^6$--$10^7$~yr after the MISO flyby), the perturbation
wave formed by secular planetary interactions propagates from the
outer Solar system to its inner zone.}

\bigskip

\noindent Keywords: celestial mechanics -- planets and satellites:
dynamical evolution and stability -- exoplanets -- ISM: individual
objects: 1I/'Oumuamua -- ISM: individual objects: 2I/Borisov

\vspace{8mm}

\section*{Introduction}

A free-floating planet (FFP) is understood as an interstellar
object of planetary mass; in other words, it is a planet that is
not gravitationally bound to any star. According to modern
estimates, in our Galaxy, the number of FFPs with Jovian and
greater masses should exceed the number of main-sequence stars by
at least several times (\citealt{Mroz_etal17}). Given the age and
number of typical stars in the Milky Way, this means that a close
encounter between a star and an FFP should not be an exotic
phenomenon. Indeed, according to \cite{GR18}, approximately $1\%$
of the number of stars with a mass less than two Solar masses
experience a temporary capture of a FFP during their lifetime;
therefore, the star fraction that experience an encounter with
FFPs should be considerably higher. The study of interactions of
planetary systems with such objects is of great interest, since
they directly relate to the problem of stability and long-term
evolution and survival of planetary systems.

The FFPs and brown dwarfs (BDs) are regarded as objects of
planetary and substellar masses, respectively. The upper mass
limit for a planet is $\sim$13~$\mjup$ (Jupiter masses). With a
larger mass value, deuterium ignites in the core and the planet
passes into the BD category (\citealt{LdEL22}); the upper mass
limit for BD is $\sim 75 \mjup$.

In this paper, we investigate the influence of flybys of massive
interstellar objects (MISOs), such as FFPs or BDs, on the
immediate and long-term dynamical evolution of the Solar system.

As for higher-mass MISOs (i.e., stars), encounters with them has
been already studied, mostly by numerical-analytical means, in
various settings in application both to our Solar system
(\citealt{MDH11, LA15, DKC19, SCS20, ZBA20}) and model planetary
systems (\citealt{LA98, ZT04, PZJ15, ZhK15, CKZ17}). For our Solar
system, it was concluded that encounters with stars that lead to
immediate destabilisation must be too rare to occur on the
timescales less than tens of gigayears.\footnote{By the
``destabilisation'' of a planetary system we imply either
immediate or time-delayed loss (caused by the interloper flyby) of
any of the system's planets.} \cite{ZT04} and \cite{BR2022}
considered long-term consequences of such encounters; they
outlined that, due to secular perturbations among planets, the
post-flyby destabilisation may reveal itself on timescales much
longer than orbital ones, after millions of years elapsed.

Apart from the stellar hazard, a potentially hazardous role of
less massive free-floating objects should not be underestimated.
According to modern concepts (see, e.~g., \citealt{Tay99,
Morbi02}, and references therein), the planets in any planetary
system are formed of relatively small building blocks ---
planetary embryos, Mercury-sized or even Mars-sized; many of them
escape from the system at its early evolution stage, when it is
violently unstable. Mercury may be nothing but a last such embryo
survived in the Solar system; however, it is also expected to
escape on a gigayear timescale; see \cite{LG09}, and references
therein. During early stages of evolution of planetary systems,
collisions of the embryos with newly-formed planets may result in
forming binary rocky planets (such as the Earth--Moon system) and
in occurrence of large obliquities of rotation axes of giant
planets (such as that of Uranus); see \cite{Saf66, ChWB96, ChW98,
ACL99, Morbi02}. As soon as such ``post-collision'' phenomena seem
to be actually present in our Solar system, one may speculate that
a huge number of embryos may indeed escape from evolving planetary
systems and form a population of interstellar objects, apart from
the large-sized FFPs that are directly observed now in various
surveys. A relevant estimate of the concentration of stars in the
Solar neighbourhood is $\approx 0.14$~pc$^{-3}$ \citep{BR2022}.
Assuming that the stars typically possess planetary systems, the
concentration of free-floating embryos, resulting from the
formation processes, could be in orders of magnitude greater.

FFPs may originate not only from planetary systems with host
stars, but also in interstellar space, as a result of
gravitational collapse of interstellar gas blobs; \cite{GR18}
point out that various formation mechanisms may provide the FFP
concentration in the Galactic Thin Disc in the range from $0.24$
pc$^{-3}$ to $200$ pc$^{-3}$.

Certain observational constraints on the flux of interstellar
objects (ISOs) in our Galactic neighbourhood exist; they can be
used to estimate the flux in dependence on the ISO's size.
According to \cite[Figure~19]{JS22}, the flux seems to follow a
$\propto R^{-3}$ object's radius distribution. If valid, this law
predicts that on average the Solar system can be crossed per year
by more than a dozen interstellar objects with $R \sim$1~km, and,
per a billion years, by more than a dozen objects with $R
\sim$1000~km.

\begin{figure}[!ht]
\begin{center}
\includegraphics[scale=0.4]{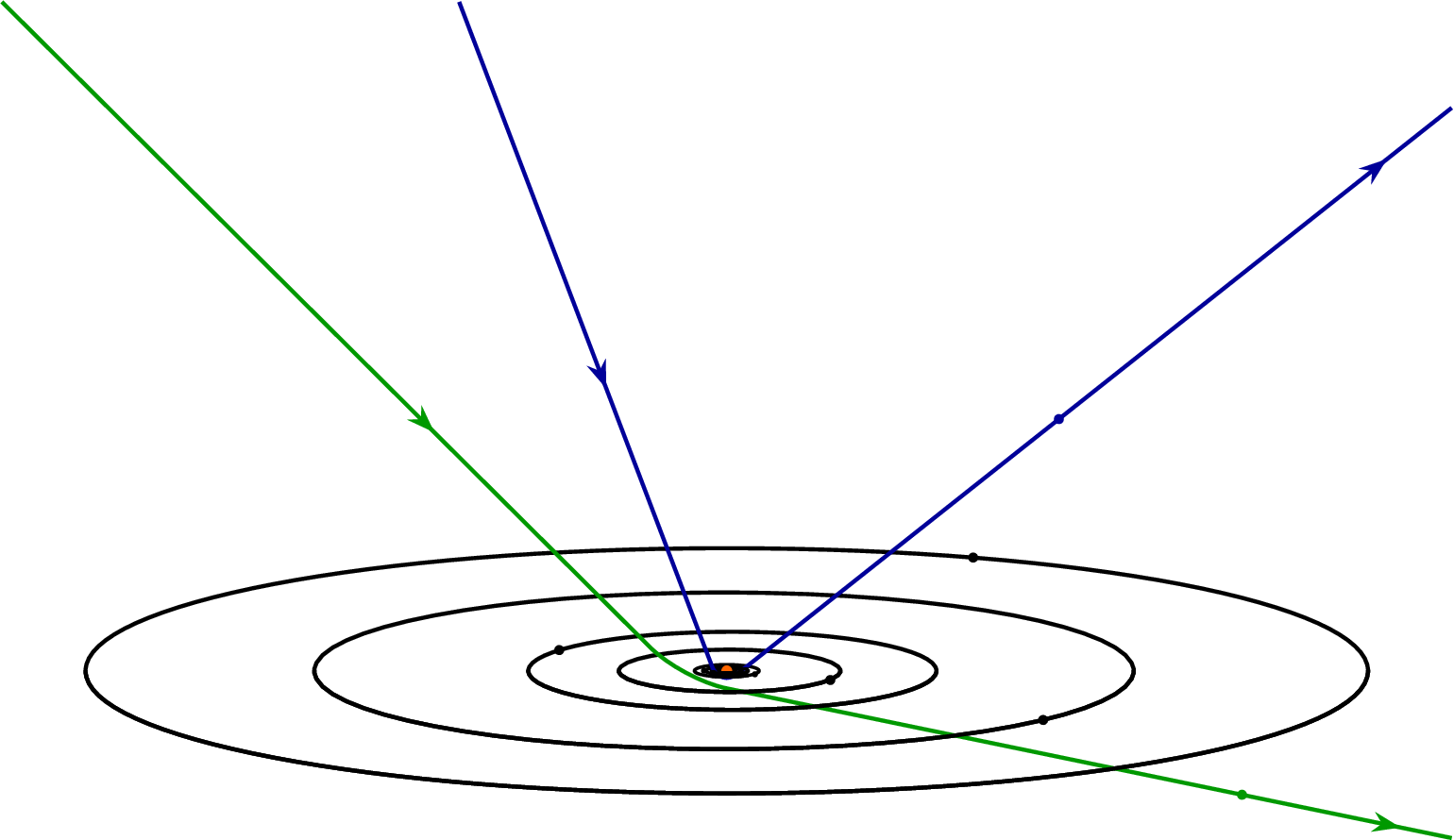}
\end{center}
\caption{A sketch of 1I/'Oumuamua (blue) and 2I/Borisov (green)
flybys. The interlopers follow hyperbolic orbits passing through
the inner regions of the Solar system. In this scheme, the
outermost planetary orbit is Neptune's.}
\label{fig1}
\end{figure}

In summary, the chance of encountering an FFP or BD by the Solar
system is definitely non-zero; this makes a study of consequences
of such an event actual. For such a study, the choice of not only
the MISO mass, but also of its orbit initial conditions, can be
broad. Here, we limit ourselves to considering two nominal MISO
approach trajectories. Namely, for the approach orbits we choose
the hyperbolic orbits of real interstellar objects 1I/'Oumuamua
and 2I/Borisov, which visited the Solar system in 2017 and 2019,
respectively (\citealt{Bannister_etal17, BL18, JL19}); henceforth
we call them orbit~I and orbit~II, respectively. We assume them to
be typical, or representative, for interstellar swarms of matter.
These orbits are specific in that they intersect the inner zone of
the Solar system; see Fig.~\ref{fig1}.

For the both approach orbits, we implement problem settings that
differ only in the MISO mass. Based on the results of our
integrations, we estimate the degree of influence of the MISO
flyby on the immediate and long-term orbital dynamics of the Solar
system.

\section*{The model set-up}

The performed numerical experiments are basically of similar
methodology kind. For each of the adopted approach orbits I and
II, a representative set of the MISO mass values is considered;
namely, the masses are taken in the ranges typical for FFPs and
BDs. The total number of numerical experiments for each of the
approach orbits is rather large ($\sim$2000), as the mass is
varied in small steps; see Table~\ref{tab1}.

\begin{table}[!ht]
\centering \hspace*{-2.5cm}\begin{tabular}{|c|c|c|c|c|} \hline
MISO type & Mass range, $\mjup$ & Step in mass, $\mjup$  &
\begin{tabular}{c} Threshold \\
                      distance $\rho$, AU
    \end{tabular}                                        &
\begin{tabular}{c} Integration time \\
                      interval $\tau$, yr
    \end{tabular}                                               \\
\hline
FFP & $0$\,--\,$13$ & $0.01$ & $1.2 \cdot 10^4$ & $5 \cdot 10^6$  \\
\hline
BD & $13$\,--\,$45$ & $0.05$ & $6 \cdot 10^4$   & $2 \cdot 10^6$  \\
\hline
\end{tabular}
\caption{The adopted values of the MISO mass, and quantities
$\rho$ and $\tau$ characterizing, respectively, the maximum
interaction distance and the integration time interval; see text
for details.}
\label{tab1}
\end{table}

The gravitational interaction of the MISO with the Sun and eight
major Solar system planets (from Mercury to Neptune) is
considered. All ten bodies are regarded as gravitating points.
Note that relativistic effects are not taken into account, since,
according to \cite{LG09} (see also \citealt{BR2023}), they
influence the Solar system stability on gigayear timescales
(diminishing chances for ejection of Mercury), whereas the
timescales of our simulations do not exceed 5 million years.

At the initial time moment $T_0$, the MISO moves at a distance
$\rho$ from the Sun; it approaches the Solar system along a
hyperbolic heliocentric orbit. The total configuration, consisting
of ten moving points, is integrated, until the MISO, after passing
the perihelion of its orbit, is away from the Sun by the same
``threshold'' distance $\rho$. The MISO is then ``switched off''
(removed from the integrated configuration), but the planetary
system is being integrated further on. The integration is stopped
when time, counted from the initial time epoch $T_0$, becomes
equal to a fixed large constant $\tau$. The adopted values of
$\rho$ and $\tau$ are given in Table~\ref{tab1}. If any of the
Solar system planets eventually escapes at a time moment less than
$\tau$, the integration is also stopped.

The threshold distance $\rho$ and the maximum integration time
$\tau$ are determined as follows. The value of $\rho$ should be
large enough so that the gravitational influence of the MISO on
the Solar system were negligible. \cite{VST12} and \cite{GR18}
modeled interaction between a Jupiter-mass MISO and the
Sun--Jupiter binary system, and they set the initial MISO
heliocentric distance equal to $40$ sizes of the binary. Here we
use a much greater (by an order of magnitude) value of $\rho$: for
FFP-type and BD-type MISOs we respectively set $\rho = 12000$~AU
and 60000~AU (i.e., up to about one light-year).

In orbits I and II, it takes respectively $\sim$4 and $\sim$20
thousand years for the MISO to cover its trajectory completely
from its initial point up to the point of its removal from the
computation.

\begin{figure}[!ht]
\begin{center}
\includegraphics[scale=.75]{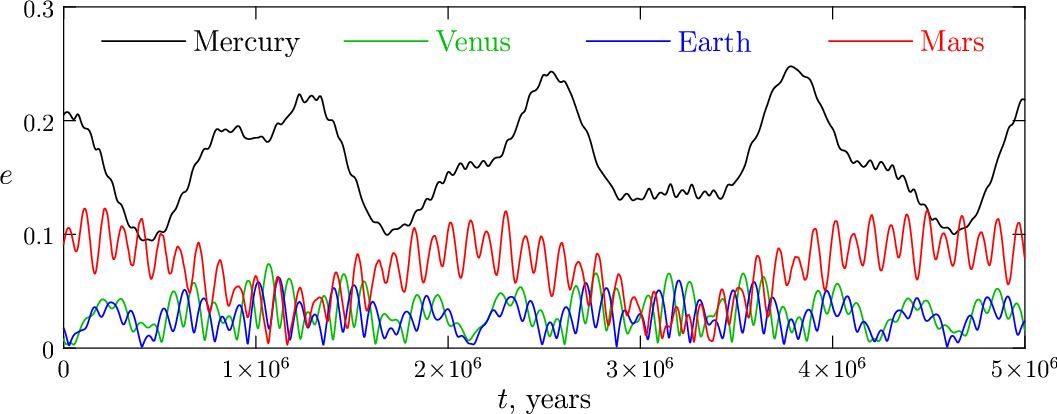}
\end{center}
\caption{Long-term evolution of eccentricities of the terrestrial
planets over 5~Myr in the absence of MISO flybys. The time is
counted from the present epoch.}
\label{fig2}
\end{figure}

We aim to study not only the immediate but also the long-term
effect of MISO flybys on the Solar system stability. To this end,
it is reasonable to choose the integration time interval $\tau$ to
be longer than, or at least similar to, the timescales of
intrinsic long-term variations of orbital elements of the Solar
system planets. Those were thoroughly explored and are well-known;
see, e.~g., \cite{Laskar96, MD99, ML21}. According to \cite{ML21},
the long-term variations of eccentricities and inclinations of the
inner planets occur on the timescale of $10^5$--$10^6$~yr; see
also our Fig.~\ref{fig2} demonstrating this behaviour on the 5~Myr
interval in the future. According to Fig.~2 in \cite{ML21} and
Fig.~9 in \cite{Laskar96}, the orbital eccentricity of Mars
fluctuates on the timescale $\sim 2.5$~Myr. In our numerical
experiments with planet-mass MISOs, we take $\tau$ to be twice
this value, namely 5~Myr. For substellar-mass MISO flybys, $\tau$
can be more than halved, because in this case the Solar system
dynamics becomes significantly unstable, as we will see further
on, on a timescale as small as 1--2~Myr.

\begin{table}[!ht]
\centering
\begin{tabular}{|c|c|c|c|c|c|c|c|}
\hline Orbit & $a$, AU & $e$ & $i$,$^\circ$ &
$T_{\mathrm{pc}}$, JD & $g$,$^\circ$  & $\Omega$,$^\circ$ &  $T$, JD \\
\hline
I                       & $-1.272$  & $1.201$       &  $122.742$    &
$2458006.007$           & $241.811$ & $24.597$      &  $2458080.5$  \\
\hline
II                      & $-0.852$  & $3.356$       &  $44.053$     &
$2458826.045$           & $209.124$ & $308.149$     &  $2459062.5$  \\
\hline
\end{tabular}
\caption{Osculating elements of orbit I (1I/'Oumuamua) and orbit
II (2I/Borisov). From left to right: semimajor axis, eccentricity,
inclination, pericentre epoch, pericentre argument, ascending node
longitude. The last column shows the time epochs for which the
elements are given.}
\label{tab2}
\end{table}

For the elements of orbit I and II we adopt those given in the
NASA JPL database (accessed January 16, 2023), see
Table~\ref{tab2}. The Solar and planetary masses are taken from
the same database.

We work in the heliocentric coordinate system $Oxyz$, to which the
orbital elements of the MISO and planets are assigned. First, we
calculate the planetary elements for the same time $T$ for which
the elements of a stellar intruder are given (see
Table~\ref{tab2}). Based on the elements and using standard
algorithms (see, e.~g., \citealt{Herrick71, Battin99, KT07}), we
obtain the positions and velocities of the planets and
interstellar object in the $Oxyz$ system at the time $T$. Then, we
numerically integrate this configuration backwards in time until
the massless interstellar object reaches the heliocentric distance
$\rho$.

\begin{table}[t]
\centering
\hspace*{-2.5cm}\begin{tabular}{|c|c|c|c|}
\hline
Orbit & MISO type &  $T_0$, JD  &
\begin{tabular}{c} Interaction time, yr
\end{tabular} \\
\hline
I   & FFP  &  $1671841.012$  (April 2, 135 BC)     & $4306$  \\
\hline
II  & FFP  &  $1815586.161$  (October 21, 258 AD)  & $3523$  \\
\hline
I   & BD   &  $-1474595.627$ (October 11, 8750 BC) & $21545$ \\
\hline
II  & BD   &  $-758768.012$  (August 10, 6790 BC)  & $17620$ \\
\hline
\end{tabular}
\caption{The initial epoch $T_0$ and the interaction time
(duration of the interaction phase, at which the MISO's
heliocentric distance does not exceed $\rho$).}
\label{tab3}
\end{table}

The configuration obtained in this way is considered to be the
initial one corresponding to the time epoch $T_0$. The interaction
time interval is approximately twice as long as the time elapsed
between epochs $T_0$ and $T$. The values of $T_0$ calculated in
our experiments are given in Table~\ref{tab3}.

In the course of integration, we calculate the maximum values of
the planetary eccentricities and inclinations
\begin{equation}
\label{eq1}
e_{\max}^j = \max e_j, \quad i_{\max}^j = \max i_j,
                       \qquad 1\leqslant j\leqslant8,
\end{equation}
as well as quantities
\begin{equation}
\label{eq2}
\begin{aligned}
d_{\min}^1 &= \min(a_3(1-e_3) - a_1(1+e_1)),  \\
d_{\min}^j &= \min(a_j(1-e_j) - a_{j-1}(1+e_{j-1})),
              \quad 2\leqslant j\leqslant8,
\end{aligned}
\end{equation}
measured in AU. To calculate all 24 quantities $e_{\max}^j,
i_{\max}^j, d_{\min}^j$, $1\leqslant j\leqslant8$, a time step of
5~yr is used, and the maxima and minima in the right-hand sides of
\eqref{eq1} and \eqref{eq2} are taken over the total integration
interval.

We also calculate the relative deviations of the energy integral:
$$
\varepsilon = \biggl|\frac{\mathcal E-\mathcal E_0}{\mathcal
E_0}\biggr|,
$$
where $\mathcal E_0$ and $\mathcal E$ are, respectively, the
initial and final values of the system total energy. The
calculation of $\varepsilon$ is carried out separately (1)~during
the preliminary backward integration, (2)~during the interaction
phase (beginning at $T_0$ and ending with the exclusion of the
MISO from the system), and (3)~during the further evolution of the
planetary system. Since the total energies at these three stages
are different, for the final energy deviation value we take the
largest of these three quantities.

\section*{Integrators and computations}

In all computations presented here, we use the IAS15
high-precision non-symplectic integrator implemented in the
REBOUND software package \citep{RL12, RS15}. The universal system
REBOUND, written in the general-purpose language C, contains a
large set of integrators designed to model a broad range of
celestial-mechanical and astrophysical problems.

In REBOUND, the equations of motion are integrated in Cartesian
coordinates, which should be referred to an inertial frame of
reference. To prevent a systematic drift of the bodies relative to
the origin of coordinates during the simulation (if the total
momentum of the system is non-zero), REBOUND can reduce the
equations of motion into a barycentric reference frame before
starting the integration. This feature is especially useful if the
computations are performed over long time intervals.

In our computations, the astronomical system of units (Solar mass,
mean Solar day, and astronomical unit) is adopted; the
gravitational constant is $G=k^2$, where the Gauss constant
$k=0.01720209895$.

The computing resources of the Joint Supercomputer Center of the
Russian Academy of Sciences (JSCC RAS, see \citealt{SSTB19} and
\texttt{https://www.jscc.ru/}) were used. Each MPI process ran one
instance of REBOUND with a given value of the MISO mass. The
numerical experiments typically took time from 15 to 25~hr each.

\section*{Post-flyby immediate and long-term effects}

\subsection*{Planet-mass MISOs}

Let us consider first the results obtained for planet-mass MISOs
in orbit~I.

\begin{figure}[!h]
\begin{center}
\includegraphics[scale=0.55]{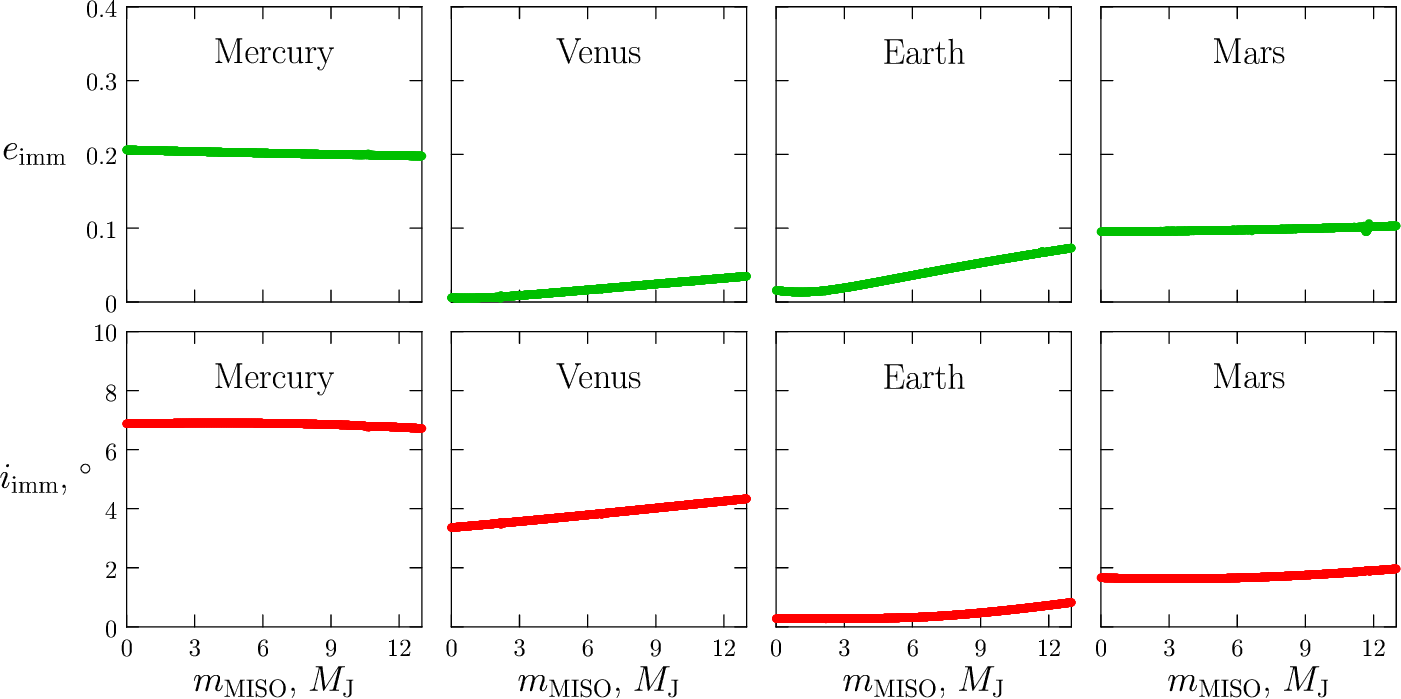}
\end{center}
\caption{The immediate eccentricities and inclinations of the
inner planets; the case of MISO in orbit~I.}
\label{fig3}
\end{figure}

The immediate (just after the flyby) eccentricities and
inclinations are presented, as functions of the MISO mass, in
Fig.~\ref{fig3} (for the inner planets) and in Fig.~\ref{fig4}
(for the outer planets). These are the values
observed at the time moment of the MISO exclusion from the system.

As follows from Figs.~\ref{fig3} and \ref{fig4}, the most prone to
immediate excitation are the orbits of the outermost planets,
Uranus and Neptune. For all other planets, the eccentricities and
inclinations are excited much less, and they demonstrate much smaller
variations with the increase of MISO's mass.

On longer timescales, the evolved orbital parameters of the inner
and outer planets are illustrated, respectively, in
Figs.~\ref{fig5} and \ref{fig6}. Namely, the graphs show, as
functions of the MISO mass, the maximum values $e_{\max}$,
$i_{\max}$, and $d_{\min}$ (defined by formulas~\eqref{eq1} and
\eqref{eq2}; for brevity, the indices, present in the definitions,
are omitted).

The function $d=q-Q$ (where, in a planetary pair, $q$ is the
pericenter distance of the outer planet, and $Q$ is the apocenter
distance of the inner planet) provides a simple estimate of the
distance between two elliptical orbits; see, e.g., \cite{MB19}.
Let us recall some basic properties of the function $d$ in the
case of non-coplanar orbits.\footnote{We consider non-coplanar
orbits, since this is the general case. For consideration of the
coplanar case see \citealt{KV99}.} If $d > 0$, then the orbits
never intersect and are always unlinked. If $d \leqslant 0$, then
there exist three possible cases: unlinking (a), linking (b), and
intersection (c). The condition $d \leqslant 0$ generally implies
cases (a) or (b); and the degenerate case (c) separates (a) and
(b). If the planetary eccentricities are not too large, the
inequality $d \leqslant 0$ indicates that the orbits are close to
crossing and that they may be linked. During the long-term
evolution, such a pair is expected to be eventually disrupted.

From Fig.~\ref{fig5}, we conclude that a planet-mass MISO flyby in
orbit~I can cause, as a secular outcome, the orbit linking in all
neighbour pairs among terrestrial planets, and thus may lead to
ejection of planets. In Fig.~\ref{fig7}, the time behaviour of
function $d$ for the inner planet pairs is shown at four
representative values of the MISO mass, taken in the vicinity of
$\miso \sim 3 \mjup$. Both of the graphs testify that close
encounters between the planets indeed become possible.

Fig.~\ref{fig8} shows how the orbital instability of the inner
planets develops after interaction with MISOs with masses greater
than Jovian. The selected $\miso$ values correspond to the most
prominent (in Fig.~\ref{fig5}) effects on the post-flyby planetary
dynamics.

For orbit~II, our numerical results in general qualitatively agree
with those discussed above for orbit~I. The immediate (just after
flyby) eccentricities and inclinations are presented in
Fig.~\ref{fig9} (for the inner planets) and in Fig.~\ref{fig10}
(for the outer planets). The evolved orbital planetary parameters
are presented in Figs.~\ref{fig11} and \ref{fig12}.

Unlike 1I/'Oumuamua, 2I/Borisov passed outside the orbit of Mars
and at a much higher velocity. According to Figs.~\ref{fig11} and
\ref{fig12}, the most noticeable long-term response to the MISO
flyby in orbit~II is exhibited by Mercury, Uranus and Neptune, but
this response is significantly weaker in comparison with the case
of orbit~I, as Figs.~\ref{fig11} and \ref{fig12} compared to
Figs.~\ref{fig5} and \ref{fig6} indicate.

\subsection*{Flyby resonant outcomes}

In Fig.~\ref{fig5}, the evolved behaviour of the inner planets
after the MISO flyby in orbit~I is definitely violent in a narrow
range of the MISO mass $\miso$ around the value of $3 \mjup$, in
comparison to its smooth response to the flybys of less or more
massive objects. What could cause such a selectivity of the
planets' reaction to the MISO mass?

\begin{figure}[!h]
\begin{center}
\includegraphics[scale=0.55]{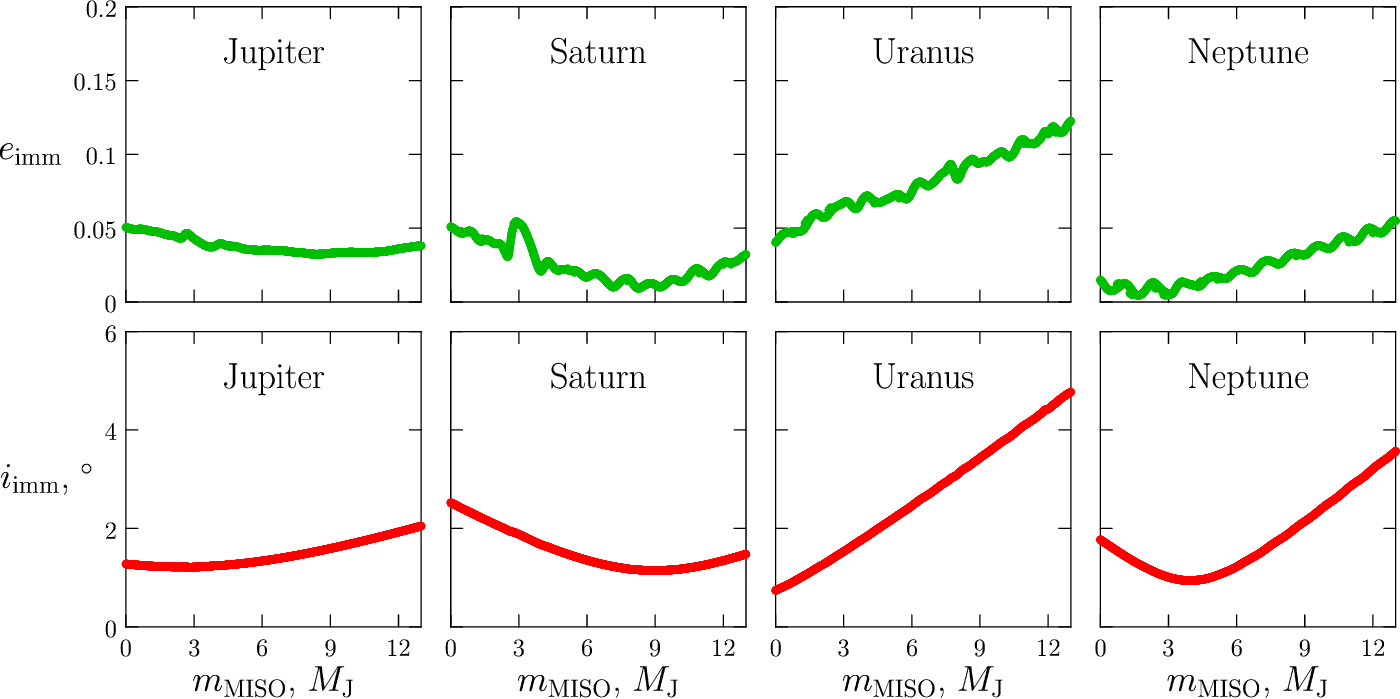}
\end{center}
\caption{The same as Fig.~\ref{fig3}, but for the outer
planets.}
\label{fig4}
\end{figure}

\begin{figure}[h]
\begin{center}
\includegraphics[scale=.55]{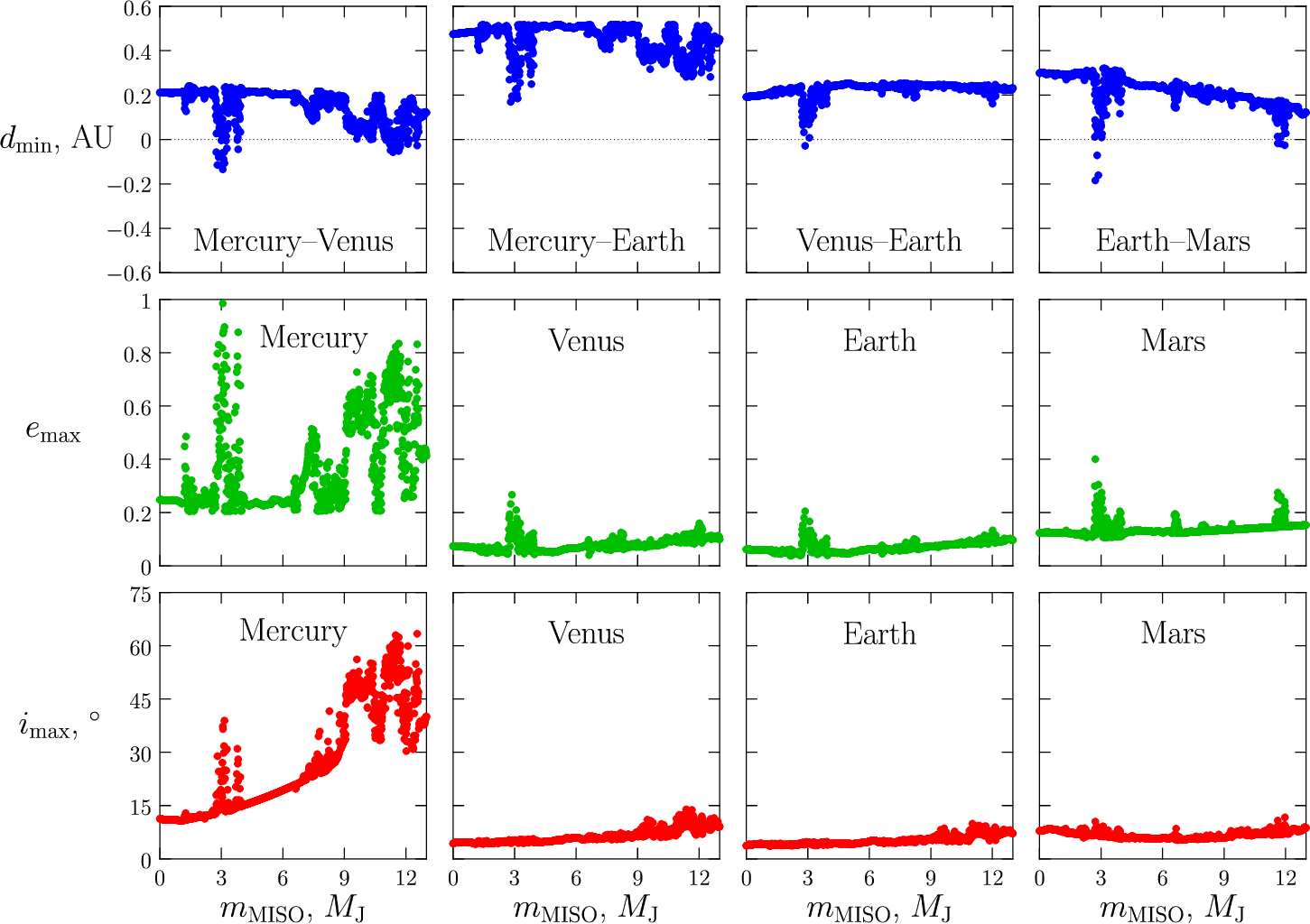}
\end{center}
\caption{Evolved orbital parameters (defined by Eqs.~\eqref{eq1}
and \eqref{eq2}) of the inner planets, as obtained in simulations
with planet-mass MISOs in orbit~I.}
\label{fig5}
\end{figure}

\begin{figure}[t]
\begin{center}
\includegraphics[scale=.55]{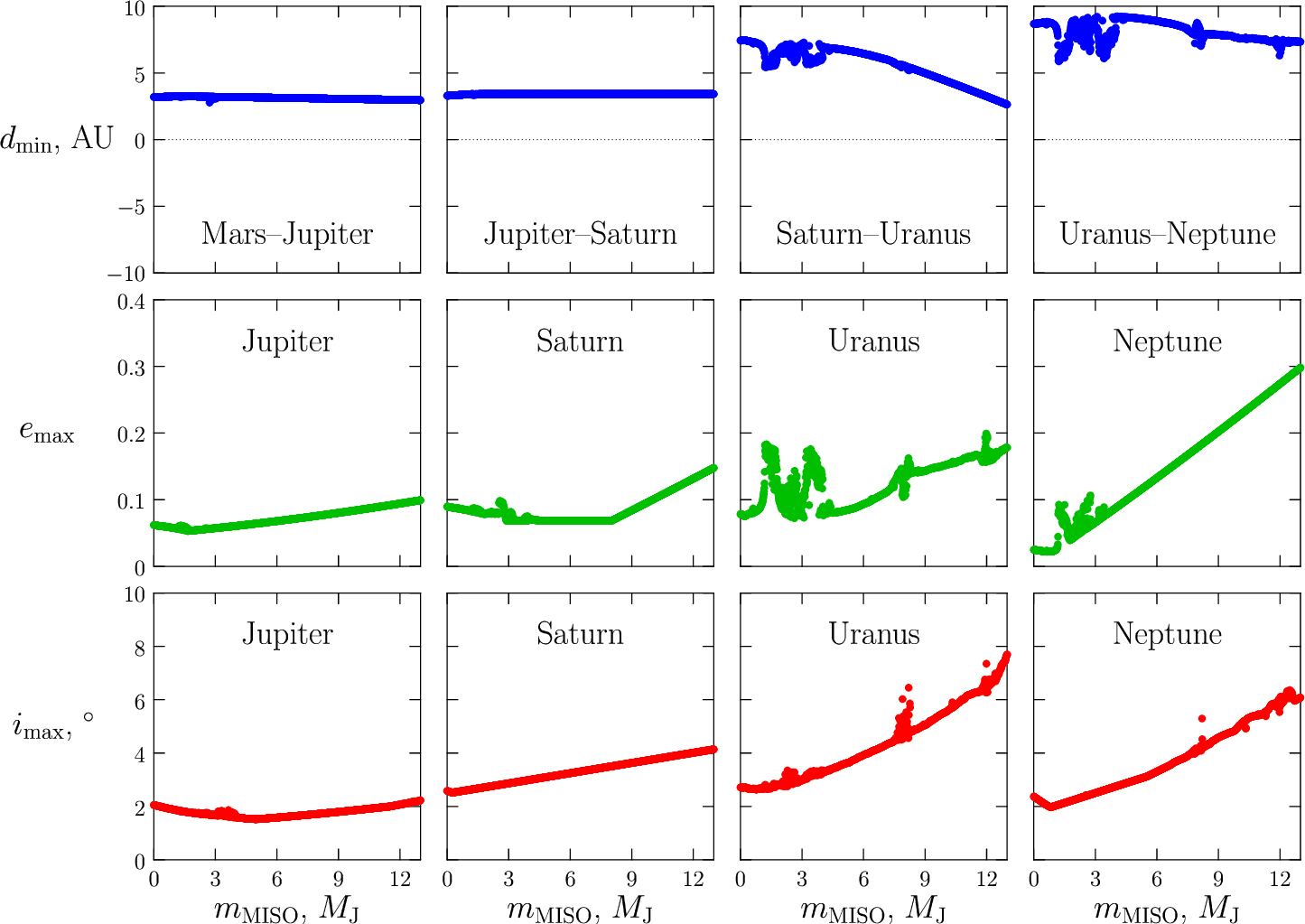}
\end{center}
\caption{The same as Fig.~\ref{fig5}, but for the outer planets.}
\label{fig6}
\end{figure}

\begin{figure}[ht]
\begin{center}
\includegraphics[scale=.75]{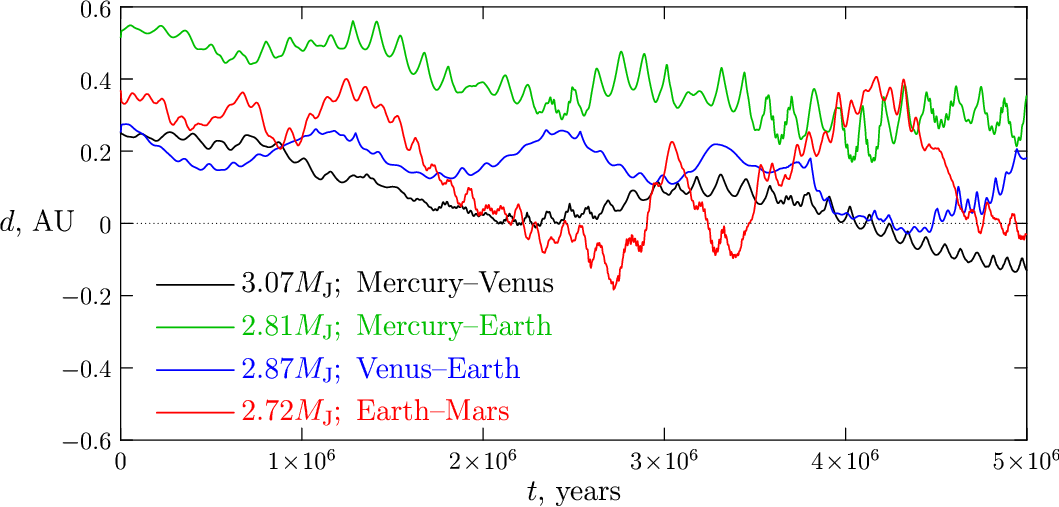}
\end{center}
\caption{The time behaviour of the parameter $d=q-Q$, where $q$
and $Q$ are the pericenter and apocenter distances, respectively,
of an outer and an inner planet, in four pairs of terrestrial
planets. The selected $\miso$ values correspond to jumps in
planetary eccentricities observed in Fig.~\ref{fig5}, in the
vicinity of $\miso \sim 3\mjup$.}
\label{fig7}
\end{figure}

\begin{figure}[h]
\begin{center}
\includegraphics[scale=.45]{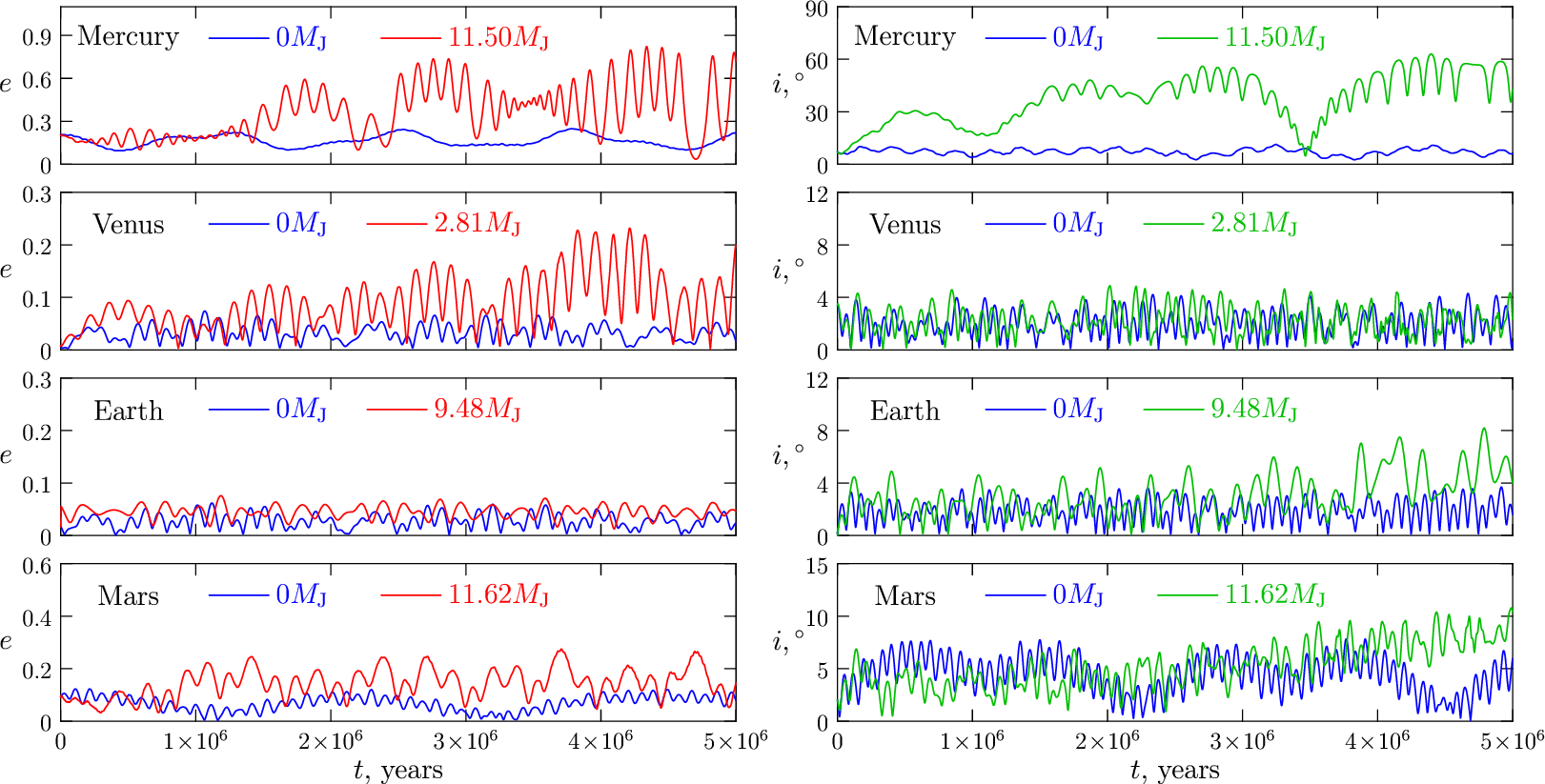}
\end{center}
\caption{Evolution of the orbital eccentricities and inclinations
of the inner planets after a MISO flyby in orbit~I. The blue
curves correspond to the unperturbed (i.e., no MISO) system. Time
is counted from the epoch $T_0$.}
\label{fig8}
\end{figure}

\begin{figure}[!h]
\begin{center}
\includegraphics[scale=0.55]{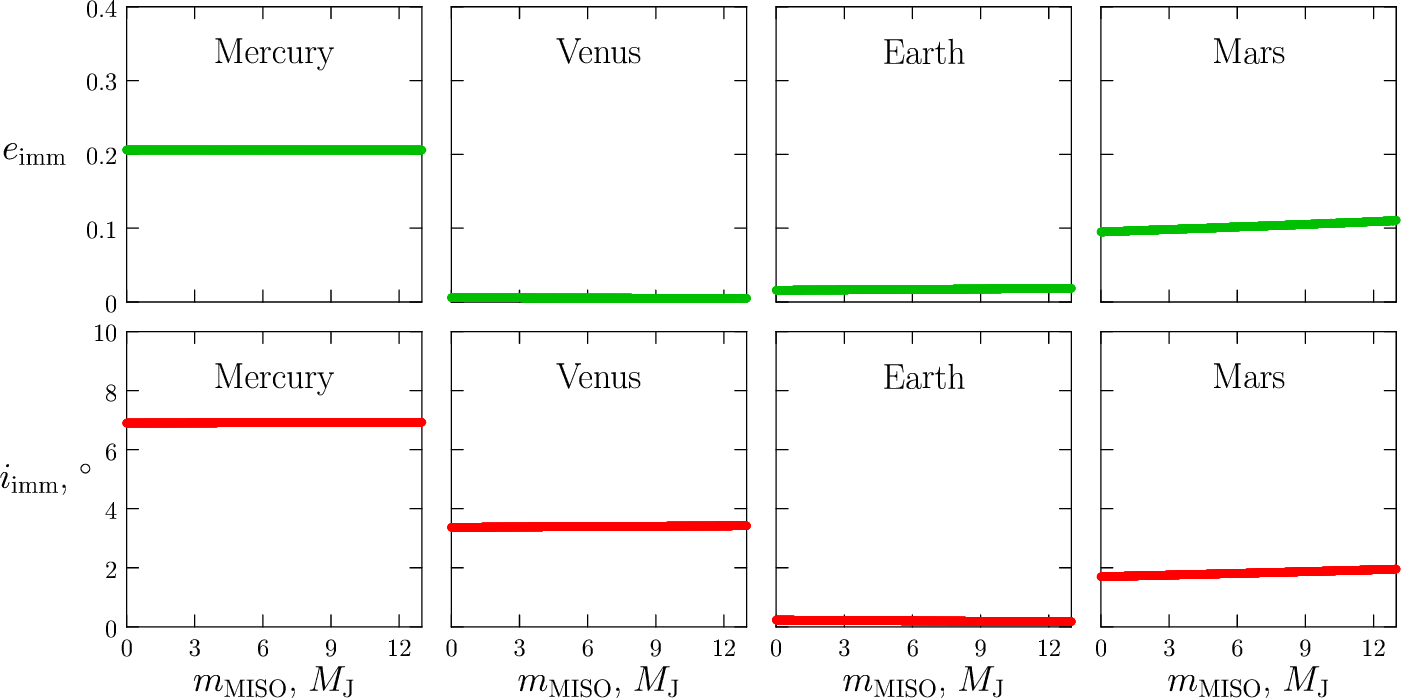}
\end{center}
\caption{The immediate eccentricities and inclinations of the
inner planets; the case of MISO in orbit~II.} \label{fig9}
\end{figure}

\begin{figure}[!h]
\begin{center}
\includegraphics[scale=0.55]{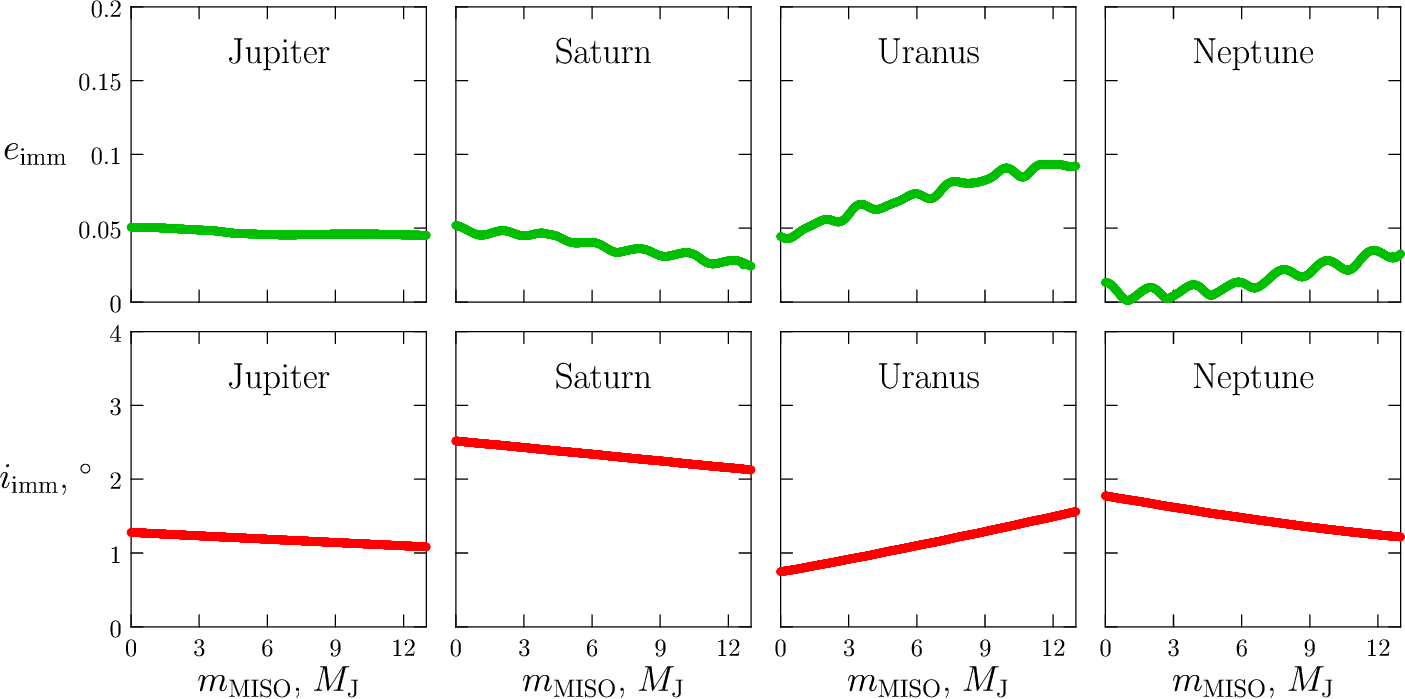}
\end{center}
\caption{The same as Fig.~\ref{fig9}, but for the outer planets.}
\label{fig10}
\end{figure}

\begin{figure}[h]
\begin{center}
\includegraphics[scale=.55]{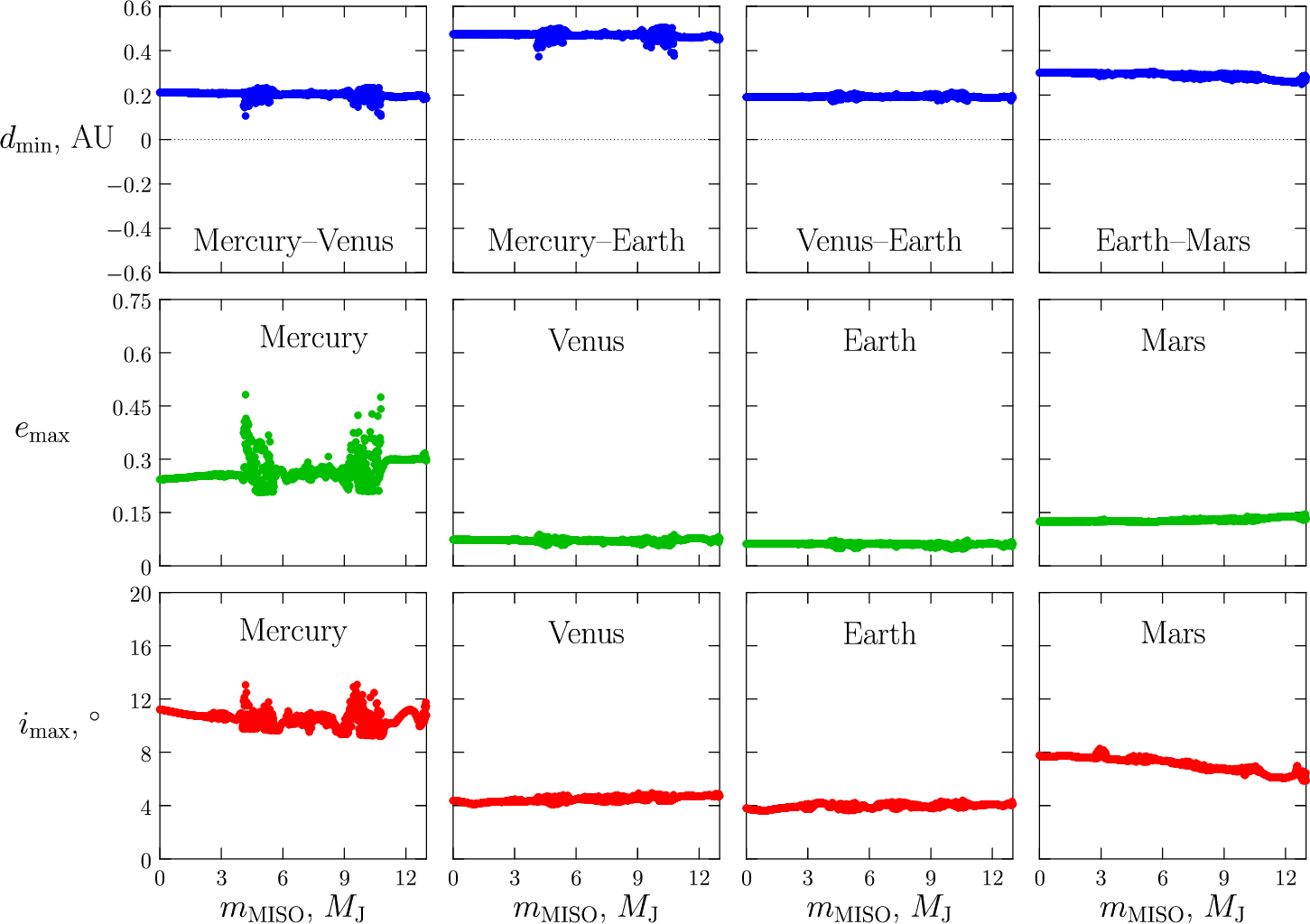}
\end{center}
\caption{Evolved orbital parameters of the inner planets, as
obtained in simulations with planet-mass MISOs in orbit~II. (I.e.,
the same as Fig.~\ref{fig5}, but for the orbit~II case.)}
\label{fig11}
\end{figure}

\begin{figure}[h]
\begin{center}
\includegraphics[scale=.55]{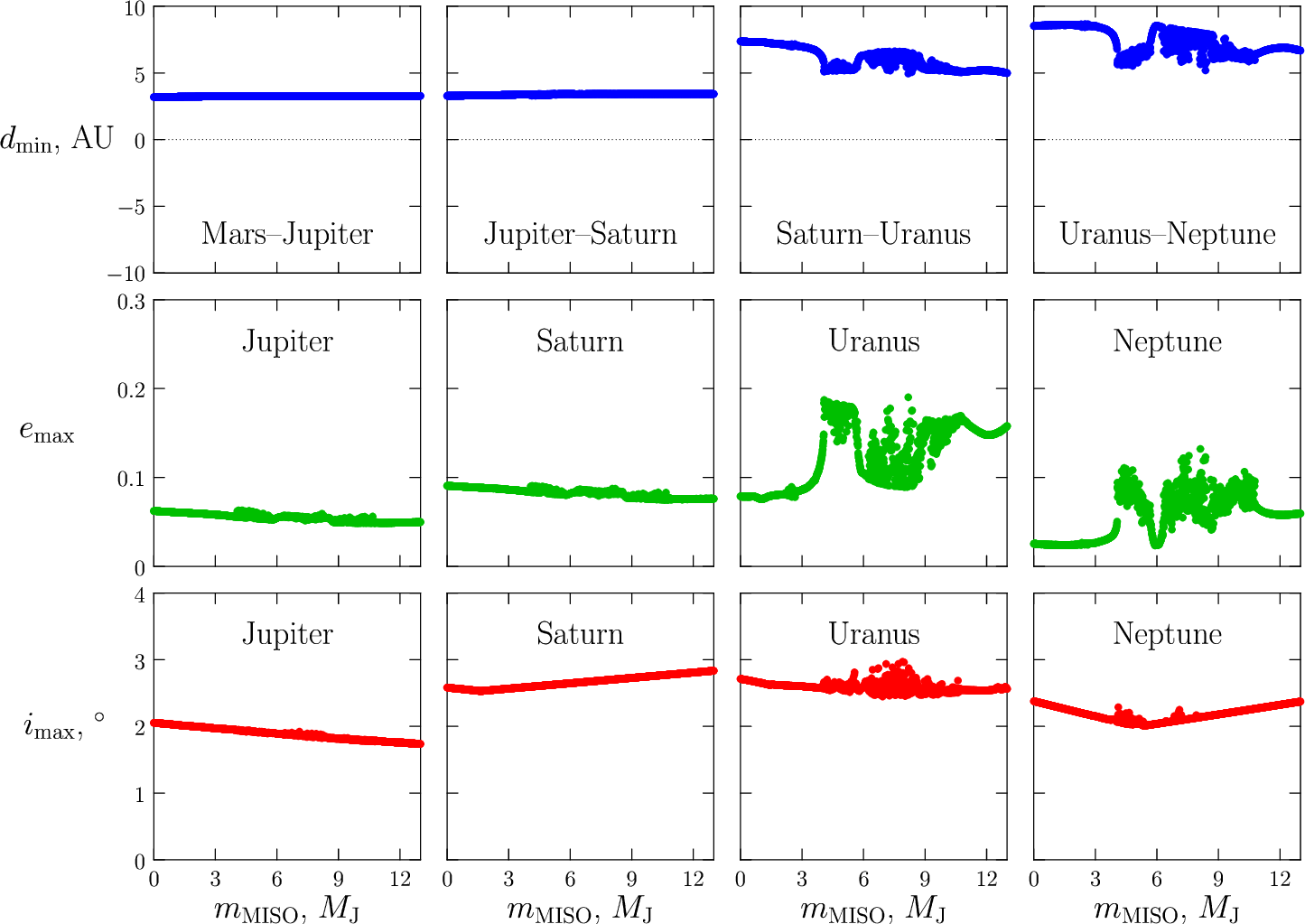}
\end{center}
\caption{The same as Fig.~\ref{fig11}, but for the outer planets.}
\label{fig12}
\end{figure}

In Fig.~\ref{fig4}, we observe an immediate (after the flyby)
increase in the eccentricity of Saturn, coherently at the same
$\miso \approx 3 \mjup$ value.\footnote{We are thankful to the
referee for turning our attention to the coherent peaks in
Figs.~\ref{fig4} and \ref{fig5}.} Naturally, the violent evolved
behaviour of the inner planets at $\miso \approx 3 \mjup$ arises
as a secular after-shock, triggered by the perturbation of
Saturn's orbit. But what could evoke this initial Saturn's
perturbation, in such a narrow specific range of $\miso$ values?

Recall that, although the flyby-caused immediate relative
perturbations of planetary semimajor axes are much less than the
reactions of eccentricities, nevertheless they are non-zero. The
flyby-induced changes in the semimajor axes may therefore,
theoretically, push the planetary system into one or another major
mean-motion resonance: of course, if the system is initially close
to the resonance, and if the flyby-caused shifts in planetary
semimajor axes have appropriate signs and absolute values to bring
the system into the resonance. If such an event happens, it will
reveal itself, in particular, in resonant excitation of
eccentricity of the least massive planet involved in the
resonance.

\begin{figure}[h]
\begin{center}
\includegraphics[scale=.75]{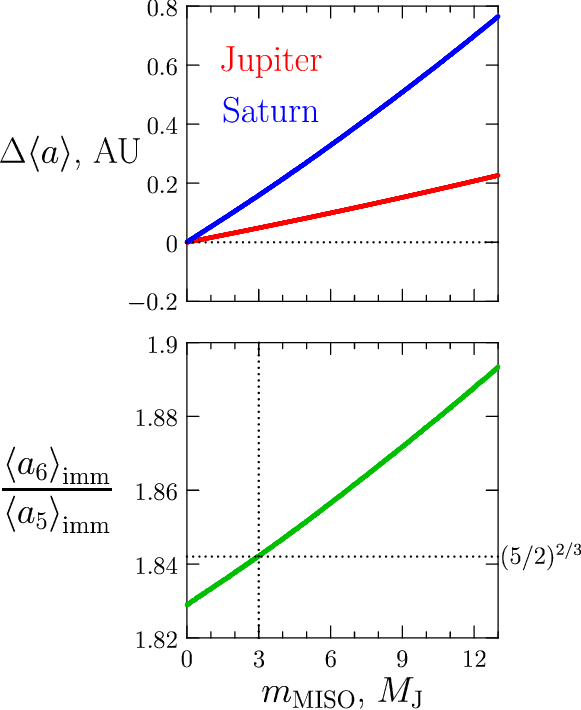}
\end{center}
\caption{The $\miso$ dependences for the immediate shifts of
Jupiter's and Saturn's semimajor axes from their nominal values
(the upper panel), and the $\miso$ dependence for the ratio of the
immediate values of Jupiter's and Saturn's semimajor axes (the
lower panel). To measure the shifts and the ratio, the semimajor
axes values are averaged just before and just after MISO's
perihelion epoch, to exclude short-periodic oscillations; see text
for details. In the lower panel, the horizontal dotted line marks
the ratio value corresponding to the mean-motion resonance 5/2;
and the vertical dotted line marks the $\miso$ value at which the
system is pushed into this resonance.} \label{fig12a}
\end{figure}

It turns out that this is just what we observe in the behaviour of
Saturn. Let us elaborate. It is well known that the
Jupiter--Saturn pair is close to the 5/2 mean-motion resonance,
but is out of it (the corresponding resonance argument does not
librate; see, e.g., \citealt{MD99}). If Saturn's semimajor axis
were by $\approx$0.3\% greater than its current value, the pair
would be in the exact resonance. In Fig.~\ref{fig12a}, we present
$\miso$ dependences for the immediate shifts of Jupiter's and
Saturn's semimajor axes from their nominal values (the upper
panel), along with a $\miso$ dependence for the ratio of the
immediate values of Jupiter's and Saturn's semimajor axes (the
lower panel). The semimajor axis shift is defined as $\Delta a =
a_\mathrm{imm} - a_\mathrm{in}$, where $a_\mathrm{in}$ and
$a_\mathrm{imm}$ are the semimajor axis values fixed,
respectively, at the start and the end of the MISO's computed
orbit. To measure the shifts and the ratio, here we have averaged
the semimajor axes values just before and just after MISO's
perihelion epoch, over time intervals of 200~yr ($\approx 10$
periods of short-periodic oscillations of Saturn's and Jupiter's
semimajor axes); the averaging is necessary to determine the
shifts and the ratio accurately enough. (In the inscriptions at
the Figure axes, the averaging is denoted by angular brackets.) In
the lower panel of Fig.~\ref{fig12a}, the horizontal dotted line
marks the ratio value corresponding to the mean-motion resonance
5/2; and the vertical dotted line marks the $\miso$ value at which
the system is pushed into this resonance. This value is just
$\approx 3 \mjup$. Thus, Fig.~\ref{fig12a} testifies that a flyby
of a MISO with mass $\approx 3 \mjup$ in orbit~I would shift
Saturn's and Jupiter's semimajor axes by amounts necessary to push
the Jupiter--Saturn pair into the 5/2 resonance.

Such an event may produce fatal, though very long-term,
consequences, as, according to \cite{ZBA20}, if Jupiter and Saturn
enter this chaotic resonance, then, over the subsequent
$\sim$10~Gyr, almost all planets (all but one) of the Solar system
are ejected. First signs of this resonance fatal secular influence
can be perhaps already identified in the evolved behaviour of
Mercury (Fig.~\ref{fig5}) and Uranus (Fig.~\ref{fig6}), just at
$\miso \approx 3 \mjup$.

Inspecting Figs.~\ref{fig4} and \ref{fig5}, along with
accomplishing numerical procedures similar to those applied above
in the Jupiter--Saturn case, reveal two more cases of
flyby-assisted resonant outcomes: at $\miso \approx 12 \mjup$, the
system is left in the 5/3 mean-motion resonance between Earth and
Mars, and, at $\miso \approx 2.4 \mjup$, in the 2/1 mean-motion
resonance between Uranus and Neptune. Considering the flyby
resonant outcomes in more detail is, however, out of scope of the
present article.

\subsection*{Substellar-mass MISOs}

An increase of the MISO mass to substellar values ($\miso > 13
\mjup$) naturally leads to a greater post-flyby
instability\footnote{By the instability (unstable state) of a
planetary system we imply here its any state that leads to its
destabilisation, defined above.} of the planetary system and to a
more rapid manifestation of the instability and disintegration.

\begin{figure}[p]
\begin{center}
\includegraphics[scale=.55]{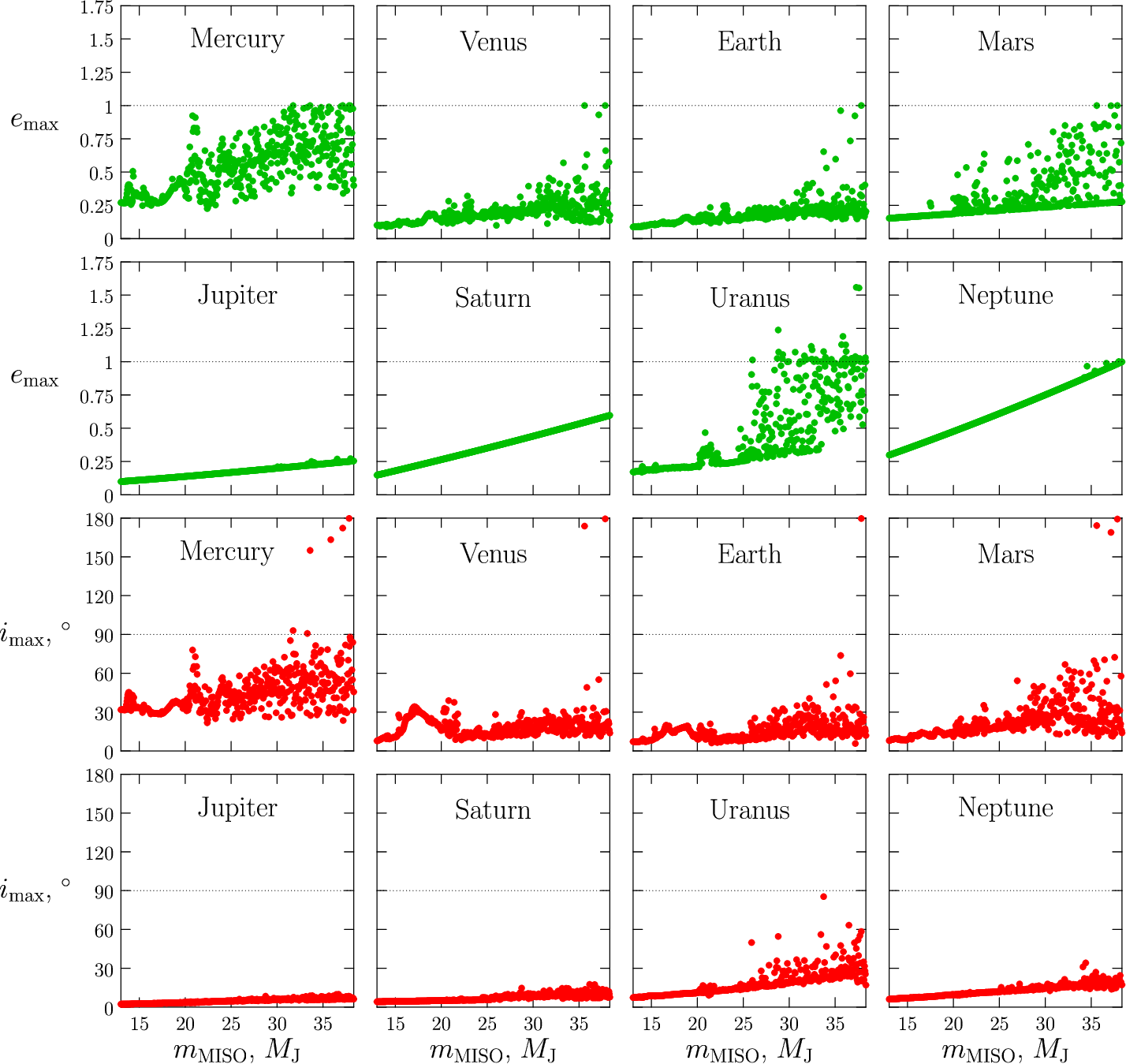}
\end{center}
\caption{Evolved orbital parameters of the Solar system planets,
as obtained in simulations with substellar-mass MISOs in orbit~I.
At $\miso\geqslant38.4\mjup$ Neptune escapes even before the MISO
is excluded, and the horizontal axis is cropped at this value.}
\label{fig13}
\end{figure}

\begin{figure}[h]
\begin{center}
\includegraphics[scale=0.55]{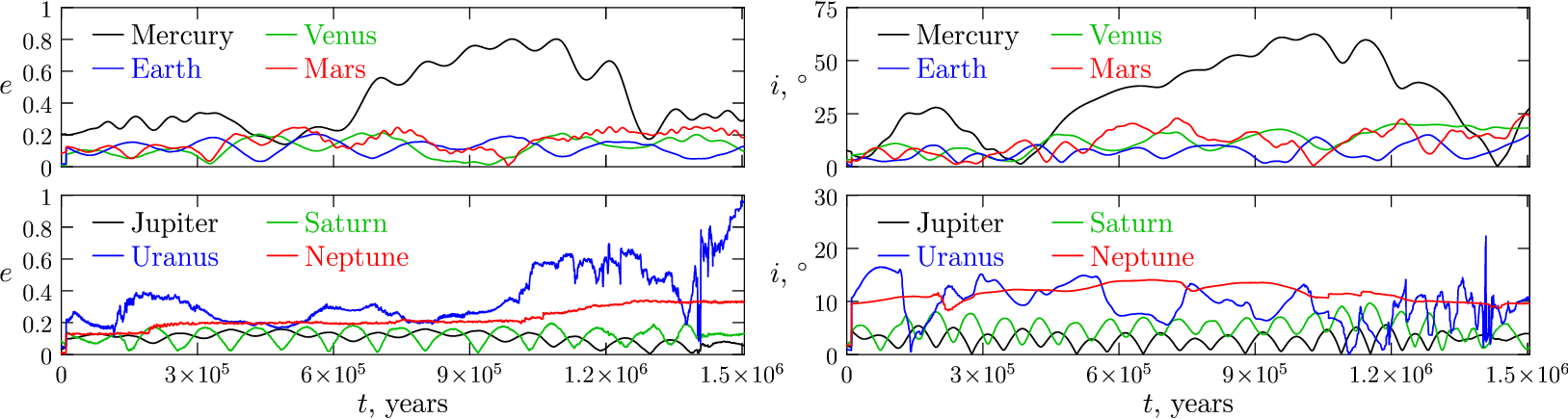}
\end{center}
\caption{Long-term evolution of the eccentricities and
inclinations of the Solar system planets before ejection of Uranus;
$\miso=29.25\mjup$, orbit~I.}
\label{fig14}
\end{figure}

Evolved orbital parameters of the Solar system planets, as
obtained in simulations with substellar-mass MISOs in orbit~I, are
illustrated in Fig.~\ref{fig13}. From this Figure, one concludes
that the flybys of a MISO with mass greater than $30\mjup$ make
ejections of Mercury and Uranus, on the long timescale, a common
occurrence. In Fig.~\ref{fig13}, the value of $\miso = 38.4\mjup$
serves as the upper boundary for the considered $\miso$ range,
since, starting from this mass value, the MISO flyby causes
Neptune to be ejected immediately; and as Neptune escapes, the
integration is stopped. In Fig.~\ref{fig14} we present, as an
example, the time behaviour of the orbital elements $e$ and $i$
for all eight planets, monitored up to Uranus' ejection;
$\miso=29.25\mjup$.

\begin{figure}[p]
\begin{center}
\includegraphics[scale=.55]{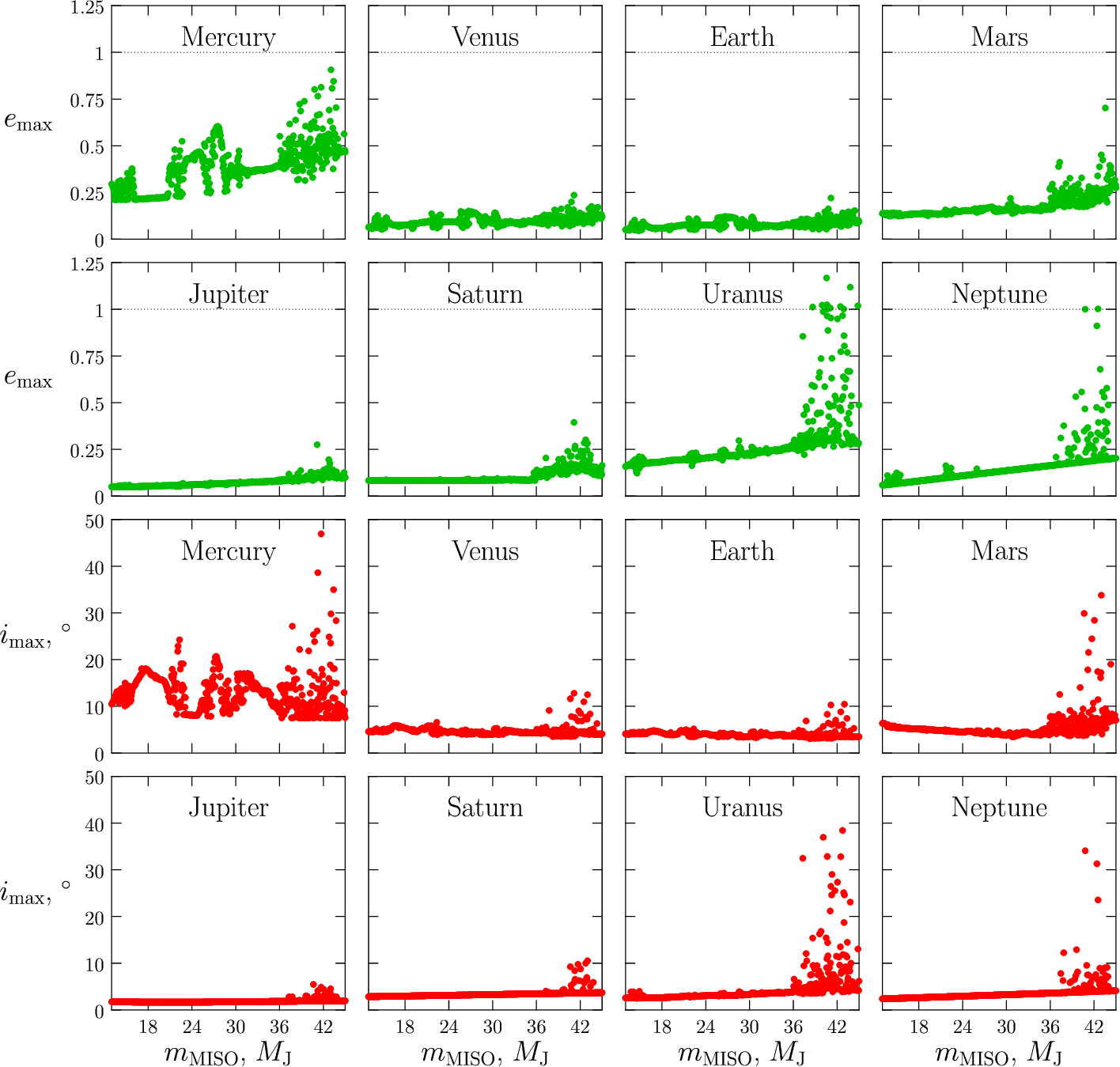}
\end{center}
\caption{The same as Fig.~\ref{fig13}, but for the case of MISO in
orbit~II.}
\label{fig15}
\end{figure}

The substellar-mass MISOs in orbit~II, as compared to those in
orbit~I, affect the planetary dynamics generally in a less
destructive way. According to Fig.~\ref{fig15}, for MISOs in
orbit~II, essentially fewer planetary ejection events are
observed. However, Uranus and Neptune are still most prone to be
ejected.

\begin{figure}[!h]
\begin{center}
\includegraphics[scale=0.2]{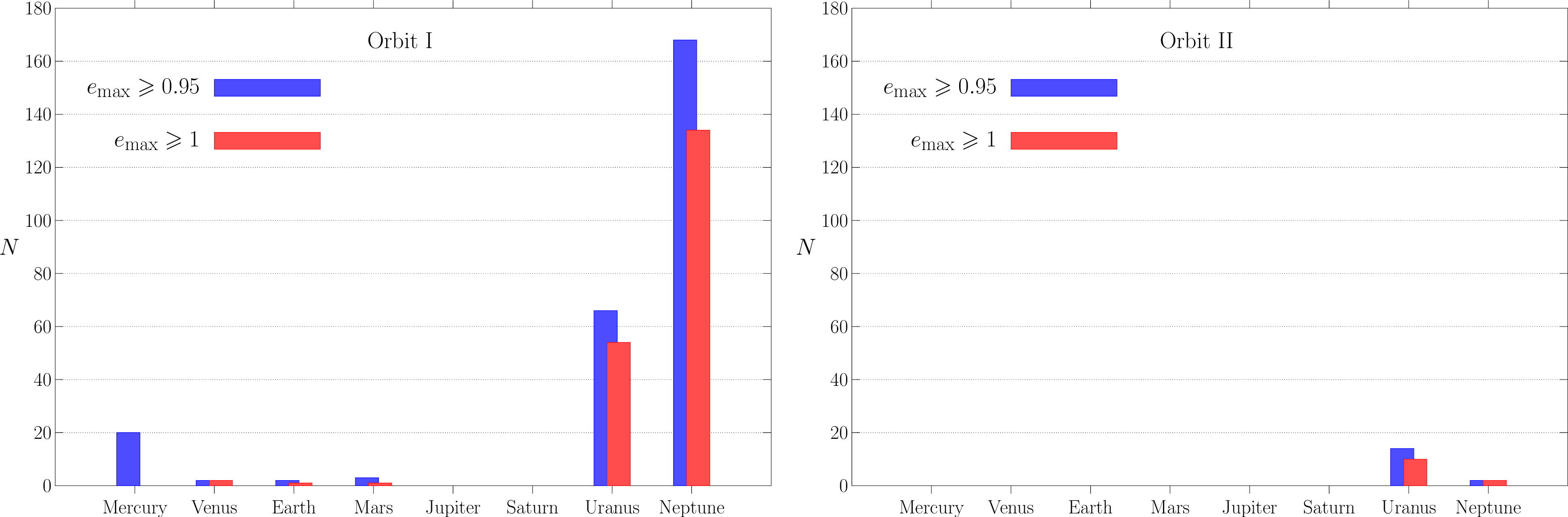}
\end{center}
\caption{The histogram (differential distribution) of all ejection
events (planet's eccentricity exceeded unity); $N$ is the number
of ejections for each planet. The histogram is built for the
simulations with substellar-mass MISOs:
$13\mjup\leqslant\miso\leqslant45\mjup$ (640 simulations in
total). Statistics obtained using a milder criterion (planet's
eccentricity exceeded $0.95$) is also shown.} \label{fig16}
\end{figure}

\begin{figure}[!h]
\begin{center}
\includegraphics[scale=0.3]{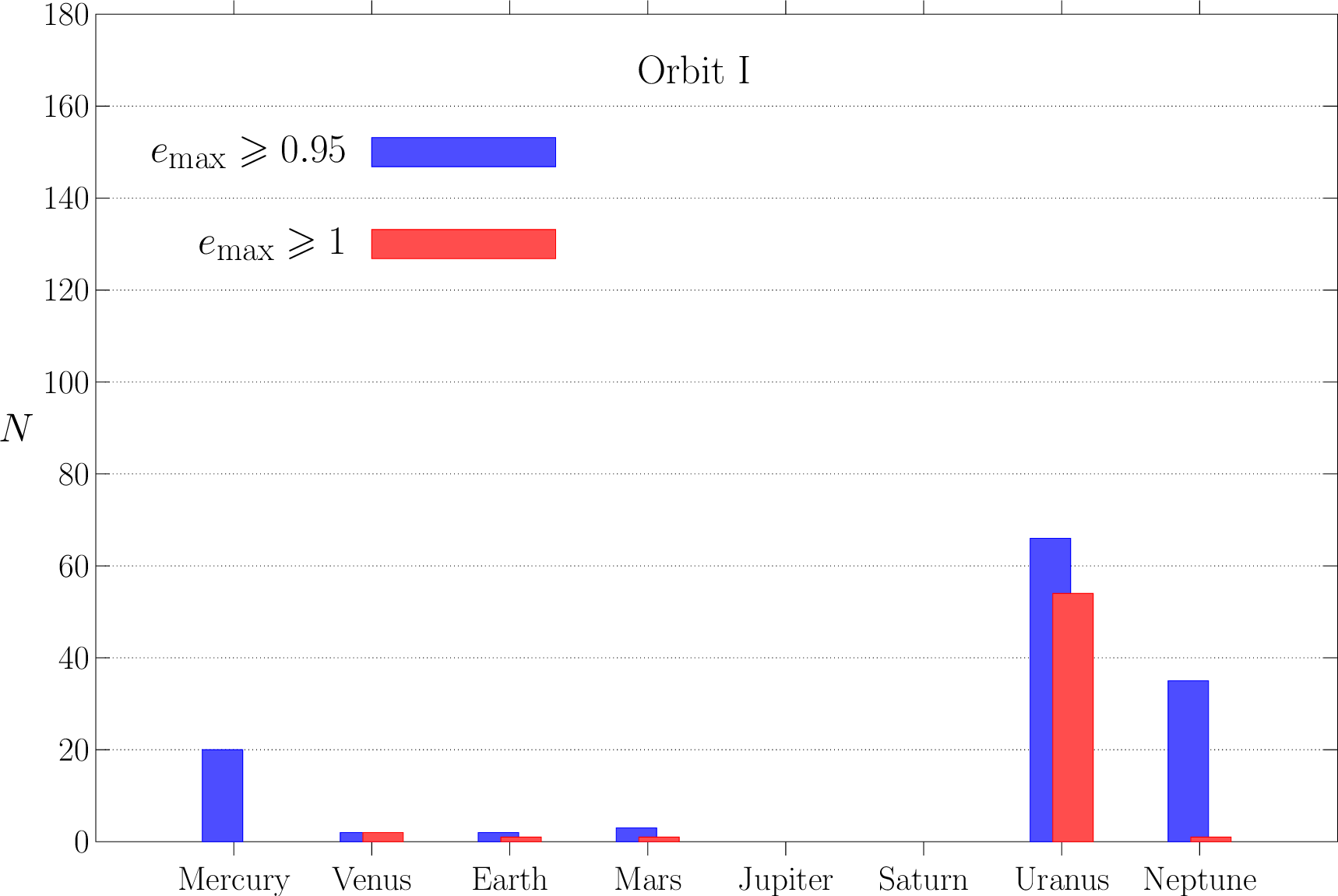}
\end{center}
\caption{The same as Fig.~\ref{fig16}, but for the orbit~I and the
MISOs with mass less than 38.4$\mjup$; the immediate regular
ejections of Neptune are thus ignored.}
\label{fig17}
\end{figure}

In Fig.~\ref{fig16}, we present a histogram (differential
distribution) of all observed ejection events. In
Fig.~\ref{fig17}, the same type distribution, but the orbit~I and
the MISOs with masses less than 38.4$\mjup$, is given; the regular
immediate ejections of Neptune are thus ignored. Two conclusions
can be obviously made: (1)~the ejections of the outermost planets
totally prevail; (2)~if the regular immediate ejections (caused by
massive-enough MISOs) of Neptune are ignored, the ejections of
Uranus (arising in the evolved system) totally prevail.

\begin{figure}[!h]
\begin{center}
\includegraphics[scale=0.7]{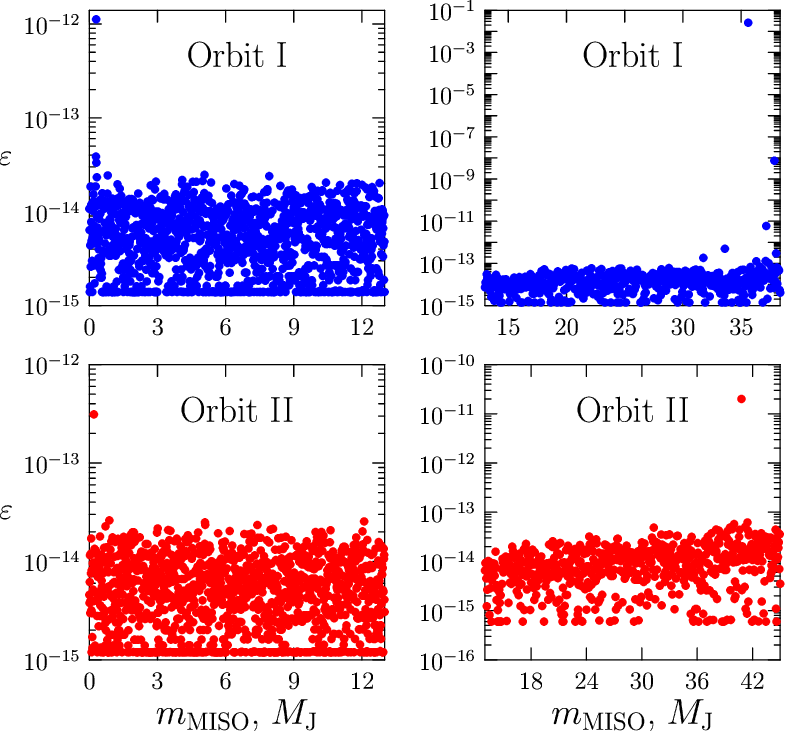}
\end{center}
\caption{The computed values of the energy deviation for the
planet-mass MISOs (left panels) and the substellar-mass MISOs
(right panels). In the top right panel there is an outlier that
occured at $\miso=35.6\mjup$. In this case, Venus was ejected at
$t \approx 3.02 \cdot 10^5$~yr, and the outlier is naturally due
to a very close encounter of Venus with another planet.}
\label{fig18}
\end{figure}

In conclusion to this Section, in Fig.~\ref{fig18} we illustrate
the energy variations as calculated at the end of each experiment.
They are negligible, thereby verifying the reliability of the
performed numerical integrations.

\section*{The flyby effect in its time development}

Let the MISO move with a velocity $v$ (velocity vector norm) and
impact parameter $p$ relative to the Sun; by $v_{\infty}$ and
$v_\mathrm{p}$ we denote the velocity $v$ values at infinity and
at pericentre, and by $q$ the pericentric distance. If the MISO
moves fast enough, i.e., the orbital periods of the planets are
much greater than the characteristic timescale of the perturbation
$\tau_\mathrm{pert}$, then the encounter can be considered in the
so-called {\it impulse approximation}; see, e.g.,
\cite{ZT04,SGHL09}.The quantity $\tau_\mathrm{pert}$ can be
defined either as $\tau_\mathrm{pert}=p/v_{\infty}$ \citep{ZT04}
or $\tau_\mathrm{pert}=q/v_\mathrm{p}$ \citep{SGHL09}. From the
data in Table~\ref{tab_tsp}, it is clear that the impulse
approximation is by far valid for all encounter cases considered
in our study.

\begin{table}[t]
\caption{Characteristic timescales of the perturbation} \centering
\begin{tabular}{|c|c|c|}
\hline Orbit & $\tau_\mathrm{pert}=p/v_{\infty}$, yr &
$\tau_\mathrm{pert}=q/v_\mathrm{p}$, yr \\
\hline
I     & $0.152$ & $0.014$ \\
\hline
II    & $0.401$ & $0.217$ \\
\hline
\end{tabular}
\label{tab_tsp}
\end{table}

Let us consider how known general relevant analytical and
numerical results, in the given approximation, can be applied to
interpret our numerical data. The effect of a passing massive body
on a stellar binary was considered in numerical simulations in the
field of stellar dynamics in several fundamental works, in
particular in \cite{Hills84} and \cite{VK05}, for either fast and
slow flybys. Here we concentrate on the changes in eccentricity,
as the relative changes in semimajor axis are generally much
smaller, $\vert \Delta a \vert / a \sim (\Delta e)^2$; see
\cite[equation~(10.119)]{VK05}. In \cite[figure 10.13]{VK05}, the
average jump $\Delta e$ in eccentricity of a binary's orbit is
given as a function of the normalized perihelion $P = q/a$, where
$q$ is the pericentric distance of the passing object, and $a$ is
the semimajor axis of the traversed binary. To construct the
graph, the numerical-experimental data obtained by \cite{Hills84}
at various masses and initial configurations of the triple system
was used. In the graph, the binary's eccentricity average
behaviour is presented for the both cases $P > 1$ and (what is
most important for us) $P < 1$. The graph shows that $\Delta e$
always declines monotonically with $P$ at $P > 0$, including the
interval $0 < P < 1$, i.e., at inner flybys.

Therefore, if one considers the traversed planetary system as a
set of binaries formed by the host star and each of its planets,
and assumes that the planets do not differ much in mass,
the greatest immediate jump $\Delta e$ is always expected for the
largest binary. In our study, these are the orbits of Neptune and
Uranus, the outermost planets, that are expected to be most
perturbed immediately, no matter whether the MISO passes outside
the Solar system, or crosses its inner zone.

This may look contradicting a common sense, but recall that the
MISO interacts with planets not only at traversing the planetary
system, but already at large distances from the Sun, and thus the
most ``loosely bound'' planetary orbits are most affected.

However, if a planet suffered a close encounter with the MISO
(their Hill spheres overlapped), then this planet is in any case
strongly perturbed.

For the outer flybys, if MISO's mass $\miso$ is much greater than
the planetary masses $m_\mathrm{pl}$ in the encountered system,
$\miso \gg m_\mathrm{pl}$, then the excited planetary
eccentricities are known to be independent of the planetary mass
$m_\mathrm{pl}$, either in fast and slow flyby cases; see,
respectively, formulas~(16) and (28) in \cite{ZT04}. Those
formulas were derived by \cite{ZT04} for the essentially outer
flyby case, but our numerical experiments demonstrate that the
$\miso$ dependence absence holds also for inner flybys. Therefore,
if close encounters are absent, one would expect that even small
bodies (Mercury, main-belt asteroids, and non-main-belt asteroid
populations) in the inner and middle systems would ``cloudise''
(i.e., obtain high inclinations and eccentricities)
non-immediately, but with a $\sim$million-year time delay, as the
terrestrial planets do, when the ``secular after-shock'' arrives
in the inner system. Conversely, the Kuiper belt objects would
cloudise immediately, in concert with Neptune and Uranus.

Close encounters are extremely rare, and one may expect that
solely the outermost Uranus and Neptune are subject to strong
immediate perturbation. If Uranus and Neptune obtain
eccentricities high enough, they start to intersect the orbit of
far more massive Saturn. If, eventually, one of them crosses
Saturn's Hill sphere, it can be ejected from the system. The
timescale for this occurrence is easily estimated basing on the
geometric probability of such crossings. Indeed, Saturn's Hill
sphere radius is $\approx 0.41$~AU, Saturn's orbital radius is
$9.6$~AU, and Uranus's and Neptune's orbital periods are $84$ and
$165$~yr. Therefore, the encounter (with Saturn) timescale for any
of them is, in order of magnitude, $\sim 10^3$~yr. Taking into
account mutual Uranus--Neptune close encounters provides a similar
estimate, $\sim 10^3$--$10^4$~yr.

Afterwards, the perturbation is being slowly transferred inwards.
A relevant toy model for the transfer can be presented by an array
of weakly coupled oscillators (pendula), considered in courses of
mechanics as a paradigm for the energy transfer process in
oscillator systems. If the first pendulum in the array is
perturbed, then the perturbation energy is slowly transferred
along the array (and ultimately to the last pendulum), on the
timescale determined by beating frequencies between the system
eigenmodes; see, e.g., figure 2.17 and problem 2.26 in
\cite{Cra68}.

The Lagrange--Laplace solution represents the Solar system secular
dynamics in the form of a system of coupled linear oscillators,
quite analogous to the coupled pendula array. (But note that each
pendulum in such an array is coupled by ``springs'' with all other
pendula in the array, not only with its close neighbours.) The
beating arises between the eigenfrequencies of the system; for
their values see table~7.1 in \cite{MD99} or table~7.1 in
\cite{Morbi02}. The perturbation forced by the MISO on the outer
Solar system is eventually transferred, on the secular timescale,
determined by the beating frequencies, to the inner system.

A pertinent (though not rigorously analogous) example of such a
behaviour is provided by the executive toy ``Newton's cradle'';
however, one should stress that the physical process here
is different. For the Solar system, one deals with a generalized
Newton's cradle.

Let us summarize the conclusions. In the case of substellar-mass
($>$13 Jovian masses) interlopers, i.e., free-floating brown
dwarfs, the following time scenario for the flyby consequences
takes place.

-- An immediate (on the timescale $\sim 10$--100~yr) consequence
of the flyby is a sharp increase in the orbital eccentricities and
inclinations of the outermost planets, Uranus and Neptune.

-- On the intermediate timescale ($\sim 10^3$--$\sim 10^5$~yr),
Uranus (mostly) or Neptune can be ejected from the system, due to
close encounters Saturn--Uranus and Uranus--Neptune.

-- On the secular timescale ($\sim 10^6$--$10^7$~yr), the major
perturbation wave formed by the secular planetary interactions
propagates from the outer Solar system to its inner zone.

Quite unexpectedly, we see that the rendez-vous consequences do
not typically imply any rapid destabilisation of the inner Solar
system (containing rocky planets and the potential habitability
zone); but, instead, these are the outermost giant planets that
are first of all ejected. Only with a million-year time lag, the
``secular after-shock'' anyway affects the inner zone.

How the long-term stability of our planetary system could be
deteriorated on longer timescales?

Either in the full Solar system, or in the system limited to four
giant planets, the Lyapunov time $T_\mathrm{L} \approx$ 5--7
million years \citep{MH99}; i.e., the system chaos arises in the
outer (giant-planet) system, and the inner system does not
contribute much. Our Solar system is close to the 5/2
Jupiter--Saturn two-body mean-motion resonance, and also to the
7/1 Jupiter--Uranus two-body mean-motion resonance. Neither of the
two resonant arguments librate, but its combination $\phi = 3
\lambda_\mathrm{J} - 5 \lambda_\mathrm{S} - 7 \lambda_\mathrm{U}$
does librate\footnote{Here $\lambda_\mathrm{J}$,
$\lambda_\mathrm{S}$, and $\lambda_\mathrm{U}$ are the mean
longitudes of Jupiter, Saturn, and Uranus, respectively.}, i.e.,
the system resides in the 3J--5S--7U three-body resonance
\citep{MH99}. According to the D'Alembert rules, the 3J--5S--7U
resonance possesses a lot of eccentricity-type, inclination-type,
and eccentricity-inclination-type subresonances. This is just
their interaction and/or overlap may cause the giant planets'
chaotic behaviour \citep{MH99}.

The location of the actual Solar system with respect to
neighbouring chaotic resonances is conveniently illustrated in a
FLI (Fast Lyapunov Indicator) dynamical chart, constructed by
\cite{G06I} on the fine grid of the initial values of semimajor
axes $a_5$ and $a_6$ of Jupiter and Saturn, whereas other initial
conditions for all four giant planets are fixed. In this
``$a_5$--$a_6$'' plane, the most pronounced (among other present)
chaotic resonance, is the Jupiter--Saturn resonance 5/2. It forms
a broad chaotic band quite close to the actual location of the
Solar system; see figure~1f in \cite{G06I}.

If Uranus is ejected, then the 3J--5S--7U three-body resonance and
the 7/1 Jupiter--Uranus two-body resonance are both no longer
present in the system; and, if other planets are weakly enough
perturbed by the MISO flyby, one may expect that the system
becomes less chaotic, in the sense that the Lyapunov exponents
diminish. However, other planets are nevertheless perturbed by the
flyby, and their eccentricities may significantly rise (whereas
relative variations of the semimajor axes are smaller). Therefore,
the chaotic zone associated with the 5/2 Jupiter--Saturn resonance
may broaden and engulf the location of the Solar system. Moreover,
as we have seen in the previous Section, at large enough MISO mass
the Jupiter--Saturn pair can be brought into the 5/2 resonance
directly, due to flyby-caused small shifts in the planets'
semimajor axes. On entering the resonance, the system becomes much
more chaotic: the Lyapunov exponents would sharply rise. According
to \cite{ZBA20}, entering the chaotic 5/2 resonance is fatal: over
the subsequent $\sim$10~Gyr, almost all planets (all but one) are
ejected.

On the other hand, a stronger chaos (smaller Lyapunov time) does
not automatically imply more rapid disintegration of the remaining
system; indeed, interrelations between Lyapunov times and
diffusion times are often complicated and system-dependent: e.g.,
chaos can be simply ``bounded'' to near-resonance regions,
implying that the system is eternally intact; for relevant
discussions, see \cite{S20DCPS,CGS22}.

\section*{Conclusions}

We conclude that the long-term stability of the Solar system can
be violated even if the interloper is not very massive (a Jovian
mass is enough) and the object does not experience close
encounters with the planets (the latter holds in our nominal
orbits). The disintegration of the planetary system does not
appear immediately, but may take place in several million years.

In what concerns substellar-mass interlopers (free-floating brown
dwarfs), an immediate (on the timescale $\sim 10$--100~yr)
consequence of the flyby is a sharp increase in the orbital
eccentricities and inclinations of the outer planets.

On the intermediate timescale ($\sim 10^3$--$10^5$~yr), Uranus
(mostly) or Neptune can be ejected from the system, due to close
encounters with Saturn and mutual encounters.

On the secular timescale ($\sim 10^6$--$10^7$~yr), the
perturbation wave formed by the secular planetary interactions
propagates from the outer Solar system to its inner zone.

As follows from our study, the most hazardous consequences of the
MISO flyby for the habitability zone consist in (1)~the secular
changes in the orbital eccentricities of the inner planets and
(2)~the cloudisation of the main belt of asteroids. The both
factors lead to the planetary climate changes; however, they start
to act with a ``secular after-shock'' (million-year) time lag. If
the habitability zone is indeed inhabited by an advanced
civilization, it luckily has a lot of time at its disposal to find
ways to withstand.

From the data obtained, one may also infer that it is unlikely
that the Solar system, which has an age of more than four billion
years, in its past was subject to numerous encounters with objects
of giant-planet and substellar masses, because this would induce
large eccentricities and inclinations and could even lead to
ejection of the outermost planets.

On the other hand, hundreds of multiplanet exosystems are observed
nowadays. As we have seen above, in the Solar system case, Uranus
seems to be most prone to be ejected due to a MISO flyby (putting
aside the rare chance of most massive ISOs, when Neptune is
regularly ejected). Therefore, if, in a multiplanet exosystem, an
apparent absence of the ``penultimate planet'' in the observed
orbital configuration (i.e., the prolonged gap in the orbital
radii distribution, just before the outermost planet), in concert
with large planetary eccentricities, is observed, this may hint
that the system had eventually suffered a rendez-vous with a
sufficiently massive ISO.

Of course, to compile a general picture of MISO interactions with
the Solar system, covering all possible rendez-vous orbits, it is
necessary to perform a much larger amount of computations with a
broad choice of initial conditions. However, it is already clear
from the just-described results that flybys of typical FFPs and
BDs can lead to a loss of stability and relatively rapid (in
comparison with the system age) disintegration of our planetary
system.

Finally, we note that, as for the possibility of directly
observing the passage of a planet-mass MISO, it is realistic only
when the flyby takes place close enough to the Solar system. The
probability of a MISO flyby outside Neptune's orbit is of course
much greater than that of any inner flyby; but a distant flyby can
be observationally missed. For assessing the risks of MISO flybys,
observational and statistical surveys of architectures of
exoplanet systems (on the subject of revealing possible flyby
influences) could be promising. At present, expected encounter
rates for Jupiter-mass MISOs form a broad range.

\medskip

\noindent {\bf Acknowledgments.} We are most thankful to the
referee for the insightful remarks and comments. We thank
V.~Sh.~Shaidulin for help in design of figures. The computations
were carried out using the computing resources of JSCC RAS. This
work was supported in part by the Russian Science Foundation,
project 22-22-00046.

\medskip

\noindent {\bf Data availability.} The data underlying this
article will be shared on reasonable request to the corresponding
author.

\newpage

\end{document}